\documentclass[12pt]{article}
\usepackage[english]{babel}
\usepackage[T1]{fontenc}

\usepackage[a4paper,top=2cm,bottom=2cm,left=3cm,right=3cm,marginparwidth=1.75cm]{geometry}

\usepackage{amsmath}
\usepackage{mathtools}
\usepackage{graphicx}
\usepackage{setspace}
\usepackage{cite}
\usepackage{tikz}
\usepackage{caption}
\usepackage{subcaption}
\usepackage{booktabs}
\usepackage{authblk}
\usepackage{gensymb}
\usepackage{multirow}
\usepackage{csquotes}
\usepackage[colorlinks=true, allcolors=blue]{hyperref}
\usepackage{textcomp, gensymb}
\providecommand{\keywords}[1]
{
  \small	
  \textbf{\textit{Keywords---}} #1
}
\title{Ab-initio study of the structural, elastic and dislocation properties of body-centered cubic refractory HfNbTaTiZr high entropy alloy}

\author[1]{Asif Iqbal Bhatti}
\author[1]{David Tingaud}
\author[2]{M. Seydou}
\author[1]{Sylvain Queyreau\thanks{Corresponding author: sylvain.queyreau@lspm.cnrs.fr}}

\affil[1]{Université Sorbonne Nord, Laboratoire des Sciences des Procédés et des Matériaux, LSPM, CNRS, UPR 3407, 99 avenue Jean-Baptiste Clément, F-93430 Villetaneuse, France}

\affil[2]{University Paris Diderot, Sorbonne Paris Cité, ITODYS, UMR 7086 CNRS, 15 rue J.-A. de Baïf, 75205 Paris cedex 13, France}

\begin{document}
\maketitle

\begin{abstract}
{This study delves into bcc HfNbTaTiZr refractory high entropy alloys, focusing on the $\frac{1}{2}\langle111\rangle$ screw dislocation core structures. While traditional observations in pure elements often revealed compact dislocation cores, our investigation reveals variability, including instances of compact, split, or degenerate cores. Building upon the current understanding of dislocation behavior, we propose that this observed variability may be intricately linked to the presence of chemical short-range order within the alloy. The unique composition of HfNbTaTiZr, between bcc and hcp elements, introduces a dynamic interplay influencing the dislocation core structure. In light of these findings, we discuss the implications for plasticity mechanisms in high entropy alloys. The presence of varied dislocation core structures suggests a complex interplay between local phases and short-range ordering, influencing the material's response to external stresses. This challenges the conventional understanding of dislocation-mediated plasticity and opens avenues for tailoring the mechanical properties of refractory alloys through the controlled manipulation of short-range order.}
\end{abstract}

\keywords{High entropy alloys, phonon analysis, Screw dislocations, Special quasi-random structures (SQSs), Density functional theory (DFT), mixing entropy, Phase stability, body-centered cubic (bcc), disordered solid solution}
\section{Introduction}

With recent advances in the study of complex concentrated alloys (CCAs), high entropy alloys (HEAs) have shown great potential for practical applications far beyond that of traditional conventional alloys by opening the unexplored alloy phase space for material design. These alloys are defined by five or more elements in nearly equal concentration \cite{Ikeda2019a, Miracle2017a, Gao} with high configurational entropy playing an important role. Thus, they lead to the formation of a simple single-phase solid solution, such as face-centered cubic (fcc), body-centered cubic (bcc), or hexagonal close-packed (hcp). Among these alloys, fcc HEAs have been widely studied, but there are comparatively few studies on bcc HEAs, especially bcc refractory HEAs (RHEAs) \cite{Dirras2016, Senkov2018a}.

With respect to orthopedic implants, HfNbTaTiZr RHEA has shown promising results, outperforming Ti-based alloys. For decades, Ti-based alloys such as Ti$_{6}$Al$_{4}$V have been used for orthopedic and dental implants, etc., but they suffer from wear resistance due to a continuous loading condition and therefore begin to produce metallic deposits that damage bone tissue and cause infections. To overcome these limitations, RHEAs have shown hopeful possibilities and are an active area of research for biomedical applications. Non-toxic refractory elements are being considered to develop implants that are not only compatible with the tissue environment but also outperform traditional Ti-based alloys. Among a variety of refractory elements, immersion tests on HfTaTi-based HEAs have demonstrated biocompatibility for medical implants. The presence of non-toxic and non-allergenic alloying elements Nb, Ta, and Zr has shown high corrosion resistance \cite{Gurel2020}. Moreover, mechanical properties such as modulus of elasticity meet the criteria for orthopedic implants \cite{Yuan2019}.

By selecting appropriate refractory elements (a combination of bcc and hcp elements), properties such as high ductility and strength at both room temperature and high temperatures can be improved by changing the composition \cite{Manzoni2020}. However, the physical reasons remain unknown \cite{Maresca2020a, Maresca2020b}. To understand these key properties, it is necessary to study the dislocation properties and further investigate how a chemical disorder plays a crucial role in improving their properties.

In pure bcc metals, strength and ductility are known to depend on the lattice friction—the so-called Peierls mechanism—and the interaction of the dislocations. The core properties of $\frac{1}{2}\langle111\rangle$ screw dislocations at low temperatures essentially play a crucial role in their plasticity. As a result of the large lattice friction experienced by the dislocations, the screw motion is thermally activated by the so-called mechanism of Kink-Pair nucleation \cite{Caillard2003} and propagation. In the last decades, extensive atomistic studies have been carried out to compute the core structure and its effects on the mobility of screw dislocations to better understand the physics of plasticity and failure in these alloys. For example, Density Functional Theory (DFT) calculations have shown that the core structure of screw dislocations in Ta, Nb, Fe, W, and V is six-fold symmetric and non-degenerate (so-called compact), with no variation in core structure observed \cite{Weinberger2013b, Dezerald2014c, Ventelon2013}.

When moving to the recent atomistic and experimental study on the bcc medium entropy alloy MoNbTi suggests the dominance of non-screw character during plastic deformation and several paths for dislocation slip \cite{Wang2020}. A similar trend has also been observed for the ternary bcc alloy NbTiZr, where the authors have shown that by decreasing the Nb content, the core structure varies continuously along the dislocation line from non-degenerate to 3-fold symmetry \cite{Rao2019}. For medium HEAs, when the hcp content is added, the local chemical fluctuation leads to the variation of the core structure along the dislocation line. For the bcc MoNbTaW alloy \cite{Yin2021}, the core structure is exactly the opposite of what has been reported for ternary alloys \cite{Yin2021, Rao2019}. It is still not clear how, in bcc RHEAs, the local complex environment affects the core structure of screw dislocations and the corresponding Peierls potentials.

Maresca \textit{et al.} has proposed a theory to explain the strengthening mechanism in bcc HEAs. In this framework, it was proposed that the strengthening mechanism does not follow the usual kink-pair nucleation, but screw dislocations adopt the kinked structure as the most stable configuration near random alloy atoms \cite{Maresca2020a}. However, the local chemical short-range order (CSRO) and atomic-level distortions were not considered, which could serve as an argument that this could be another source of the strengthening mechanism in these bcc HEAs. On the other hand, the origin of the high strength at high temperatures for bcc HEAs was developed by considering the edge dislocation motion, which is largely ignored for most bcc systems \cite{Maresca2020b}.

In a multi-alloy system, the effect of pairings that form between atoms leads to the existence of a local CSRO \cite{Fernandez-Caballero2017a, DesignedResearch;J2018}. DFT-based Monte Carlo (MC) simulations have been successfully used to study the SRO in HEAs. Yin \textit{et al.} investigated the effect of chemical SRO on the core structure of bcc MoNbTaW RHEAs using DFT-based MC simulations. In their analysis, they showed the core energies are insensitive to the degree of chemical SRO \cite{Yin2021}. Moreover, for bcc MoNbTaVW and its subsystems: MoTaVW, MoNbVW, and CrTaTiVW alloys, DFT-based MC has predicted the existence of SRO \cite{Fernandez-Caballero2017a, Sobieraj2020}.

Experimentally, in bcc equimolar HfNbTaTiZr RHEA, extensive screw dislocations are observed. They are mostly screw and straight in character. Plastic deformation is controlled by the mobility of the screw dislocations. Uniaxial tests show high yield strength. Moreover, microstructural analysis has revealed the formation of super-jogs by the collisions of kinks at RT and the existence of strong lattice friction using TEM analysis, showing that the Peierls mechanism is active. The explanation of these mechanisms by the usual kink-pair theory has been considered, but it has not satisfactorily explained the Peierls mechanism \cite{Dirras2016, Lilensten2018, Senkov2019, Senkov2011}.

Recently, Huang \textit{et al.} developed the interatomic potential for HfNbTaTiZr alloys and used this potential to study the CSRO of HfNbTaZr. The author claims that it can be used to study CSRO in quinary systems \cite{Huang2021}. On the other hand, the DFT approach has been successfully used to understand the structural, elastic, and dislocation properties of pure systems. Our motivation for this work is to study the random atomic structure that mimics the RHEA structure, further investigate the formation energy and elastic properties, rationalize the plastic behavior by studying the core structure of the screw dislocations, and statistically investigate the effects of local chemical disorder on the dislocation core. To our knowledge, this is the first study on this system.

The remainder of the paper is organized as follows: Section 2 describes the computational techniques and methods used to generate the random cell. Section 3 presents the results and discussions, and the last section summarizes the conclusions.

\section{Methodology and Computational details}

In this paper, all calculations were performed using Density Functional Theory (DFT) as implemented in the Vienna ab initio simulation (VASP) code \cite{Kresse1993}. The electron core energy was approximated with plane wave basis projector augmented wave (PAW) pseudopotentials \cite{Kresse1999} and Perdew-Burke-Ernzerhof (PBE) generalized gradient approximation was used to approximate the exchange-correlation functional (XC)  \cite{Perdew1996}. The plane-wave cut-off energy was set to 600 eV. The first-order Methfessel-Paxton method \cite{Methfessel1989} with a smearing parameter of 0.1 eV was used. Brillouin zone integration was performed with the $\Gamma$-centered k-point grid.

\subsection*{Construction of a defect-free bcc HfNbTaTiZr supercell}

Mimicking and building the random disordered structure in a small supercell via the DFT simulation raises an additional problem stemming from the periodic boundary condition (PBC). The lattice parameter, $a_{0}$, for equiatomic bcc HfNbTaTiZr RHEA was obtained by modeling the disordered structure. The short-range correlation that exists between the periodic images needs to be considered, so several techniques have been proposed in the literature to deal with this problem \cite{Glass2006, Soven1967, Gao2017b, Song2017}. In this work, we used Monte-Carlo special quasi-random structure (MCSQS) as implemented in the Alloy Theoretic Automated Toolkit (ATAT) code \cite{VandeWalle2013, VanDeWalle2009} to obtain a random structure in a small periodic supercell. Since MCSQS is based on the Monte-Carlo (MC) approach, each SQS cell can be started independently with different random seeds, and the quality of the random supercell can be guaranteed.

For HfNbTaTiZr, a supercell of 125-atom was generated based on $5\times5\times5$ cubic primitive unit cells to account for the equimolar composition with bcc lattice symmetry. For this supercell, $\Gamma$-centered k-point grid of $2\times2\times2$ was used, and the system was allowed to relax fully, including atomic positions and lattice parameters with a conjugate gradient algorithm, and was assumed to converge until each force component acting on a given atom is less than 0.01 eV/\AA. The energy convergence test in terms of plane-wave cut-off energy and k-points was performed and found to be sufficient with an accuracy of 10 meV/atom.

To evaluate the dynamic stability of the SQS structure, we then increase the force convergence criteria to 0.01 meV/\AA~for the lowest configuration and minimize the stress tensor to 0.005 GPa. For dynamic stability analysis, we use a finite displacement method as implemented in phonopy code \cite{Togo2015} which relies on the frozen-phonon approximation within the harmonic approximation combined with VASP. Thermodynamic properties were computed using the phonopy code \cite{Togo2015}. All phonon band structures were plotted using the sumo code \cite{Ganose2018}.

For the ternary and quaternary equimolar alloy, the supercell was generated using the $3\times3\times3$ and $4\times4\times4$ bcc cubic unit cells to contain 54 and 128 atoms in a supercell, respectively.

\subsection*{Dislocation dipole supercell construction}

Another simulation involves the study of dislocations. To study this system, a dislocation dipole simulation box was set up in the desired direction to analyze the core structure of screw dislocations in the quinary alloys. The box was oriented in the $a_{1}=a_{0}[\bar{1}\bar{1}2]$, $a_{2}=a_{0}[1\bar{1}0]$, $a_{3}=a_{0}[111]/2$ direction (by using suitable scaling parameters n, p, and q for each $a_{i}$, the quadrupole arrangements of the dislocation cores can be determined). $a_{0}$ is the relaxed lattice parameter obtained from the HfNbTaTiZr supercell. Following the approach proposed by Li \textit{et al.} \cite{Lib} the size of the simulation cells can be reduced from the orthogonal to the tilted geometry by using the three-edged vector scheme: $X=7a_{1}$, $Y=3.5a_{1}+5.5a_{2}+0.5a_{3}$, $Z=2a_{3}$. With this tilted dipole geometry, a nearly square quadrupolar arrangement is maintained. Along the $[111]$ direction, the length of the supercell was doubled to minimize correlation with its periodic images. The stresses induced by inserting screw dislocations with Burger’s vectors $+$b and $-$b along the $\frac{1}{2}\langle111\rangle$ are reduced by adding the $0.5a_{3}$ component to the $Y_{}$ edge vector. The dipole cell was relaxed to the minimum energy core state using the convergence criterion on force 0.01 eV/\AA. With the settings described above, a $\Gamma$-centered k-point mesh of $1\times1\times6$ was used, and the cut-off energy for plane waves was set to 500 eV. To visualize the screw dislocation’s structure, a Differential Displacement map (DD) was used \cite{Hale2017}.

\section{Results and Discussion}
\subsection{Generation and stability of the defect-free atomic structure}

\begin{figure}[ht]
    \centering
    \includegraphics[scale=0.20]{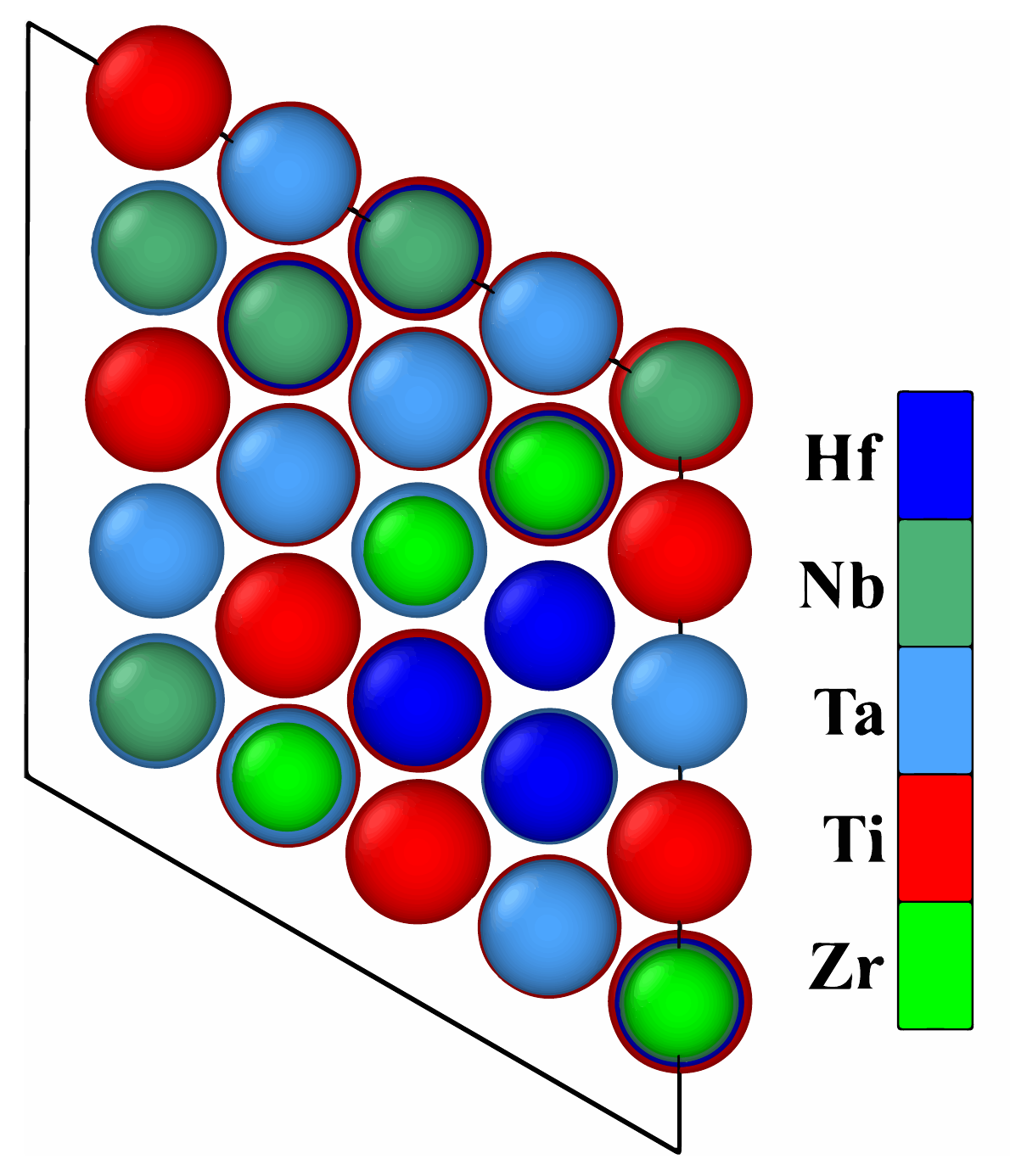}
    \caption{Construction of a random primitive supercell using the SQS approach for equimolar composition. The supercell is composed of $5\times5\times5$ times the primitive bcc unit cells. Recall that a primitive cell is half the size of a conventional bcc cell. Colored atoms are shown with constituent atoms.}
    \label{fig:bcc_SC1}
\end{figure}

Experimentally, Senkov \textit{et al.} has investigated the structure stability of HfNbTaTiZr alloy via XRD and showed that it consists of a single bcc phase \cite{Senkov2011}. Thus, to model the bcc random structure, we employed not only the nearest neighbor (NN) shells for the pairs (p) but also the triplets (t) and quadruplets (q) to minimize the correlation within the supercell \cite{VandeWalle2013}. Figure \ref{fig:bcc_SC1} shows the supercell with 125 atoms based on $5\times5\times5$ cubic primitive unit cells.

To measure the stability, we use Gibbs free energy, which is given by:

\begin{equation} \label{eq:Gibbs}
    \Delta \mathrm{G}=\Delta \mathrm{H} -\mathrm{T}\Delta \mathrm{S}^{mix} + (\mathrm{E}^{ZPE}-\mathrm{T}\Delta \mathrm{S}^{vib})
\end{equation}

In Eq. \ref{eq:Gibbs}, the mixing energy, $\Delta$H, is computed from the energy difference between the SQS supercell, E$_{SQS}$, and the most stable phase of the alloy bulk elements, E$_{i}^{bulk}$:

\begin{equation} \label{eq:formation}
    \Delta \mathrm{H} = \mathrm{E}_{SQS}-\sum_{i}\mathrm{c}_{i}\mathrm{E}_{i}^{bulk}
\end{equation}

where $c_{i}$ is the atomic fraction of the alloy component, $i$. E$_{i}^{bulk}$, the stable phase of the constituent elements. The 2nd term in Eq. \ref{eq:Gibbs} is the mixing entropic term of an ideal solid solution and is given by, 

\begin{equation} \label{eq:mixingEntropy}
    \Delta \mathrm{S}^{mix} = -\mathrm{N}k_{B}\sum_{i}\mathrm{c}_{i}ln\left(\mathrm{c}_{i}\right)
\end{equation}

Finally, the vibrational entropic term is obtained from the phonon calculations. Eq. \ref{eq:mixingEntropy} can be reduced to $\Delta \mathrm{S}^{mix} = k_{B}ln(n)$ ($k_{B}$ is the Boltzmann constant and $n$ is the number of alloy components) for equimolar composition.


To obtain the lowest possible configuration, we followed the procedure as in (\textit{Asif I. Bhatti et al. Computational material science 2024}) and performed the analysis with different (p,t,q) pairs. We showed that to model equimolar composition using MCSQS, it is enough to model random structure using (8,0,0) NN pairs. In this article, we use the same methodology and generate a 125-atom supercell corresponding to the lowest mixing energy of the structure and estimate the lattice parameter. The mixing energy corresponding to the lowest configuration we have found is 84.2 meV/atom, which is quite lower than reported by Gao \textit{et al.} 86.5 meV/atom \cite{Gao2016c}.

In Table, \ref{tab:T2} distorted lattice parameters of the SQS supercell during relaxations are shown. It can be seen that for the lowest energy configuration, a slight deviation from the cubic symmetry is observed, which can be rationalized based on the fact that the environment is composed of different atomic radii and the competition between different lattice parameters competes to obtain optimal structural stability. The average lattice, $a_{avg}$, is calculated from the average of the three distorted lattices. For the bcc SQS supercell, the average lattice parameter is 3.402 \AA $~$in agreement with the experimental value. Moreover, there is not much deviation from the experimental value, which ensures that the local fluctuations due to chemical disorder have a very weak influence on the basis vectors of the supercell.

\begin{table}[ht]
    \centering
    \caption{Calculated lattice parameters for the HfNbTaTiZr SQS supercell. The lattice parameter is reported for the conventional unit cell.}
    \label{tab:T2}
    \begin{tabular}{lcccccc}
        \toprule
         SQS& a&b&c&$\alpha^{\circ}$&$\beta^{\circ}$&$\gamma^{\circ}$\\ 
         \cmidrule{2-4}
         & \multicolumn{3}{c}{\AA} &&& \\
         \midrule
            SQS-125 [This work]	  &3.424	&3.376	&3.405	&90.1	&89.8	&89.9\\
            EXP \cite{Huang2021}  &3.404	&3.404	&3.404	&90		&90     &90\\
         \bottomrule
    \end{tabular}
\end{table}

\subsection{Structural properties and phase stability}

To rationalize our alloy's phase stability, one approach is to construct a phase diagram. We make use of the existing Materials Project database and find stability using the convex hull $(E_{Hull})$ construction. To our knowledge, we were unable to identify the decomposition products of our compound, indicating a deficiency in the DFT data for HEA, which could be crucial for training ML/DNN and potentially predicting novel properties. As a result, we relied on expensive vibrational analysis of the lowest energy structure to calculate the temperature dependence.

\begin{figure}[th]
    \captionsetup[subfigure]{labelformat=empty}
    \begin{subfigure}[b]{0.4\textwidth}
        \includegraphics[width=\textwidth]{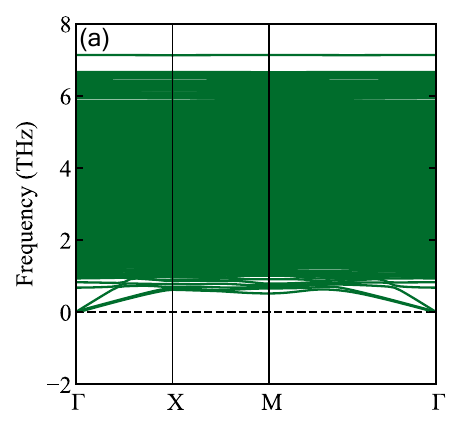}
        \caption{$~$}
    \end{subfigure}
    \begin{subfigure}[b]{0.55\textwidth}
        \includegraphics[width=\textwidth]{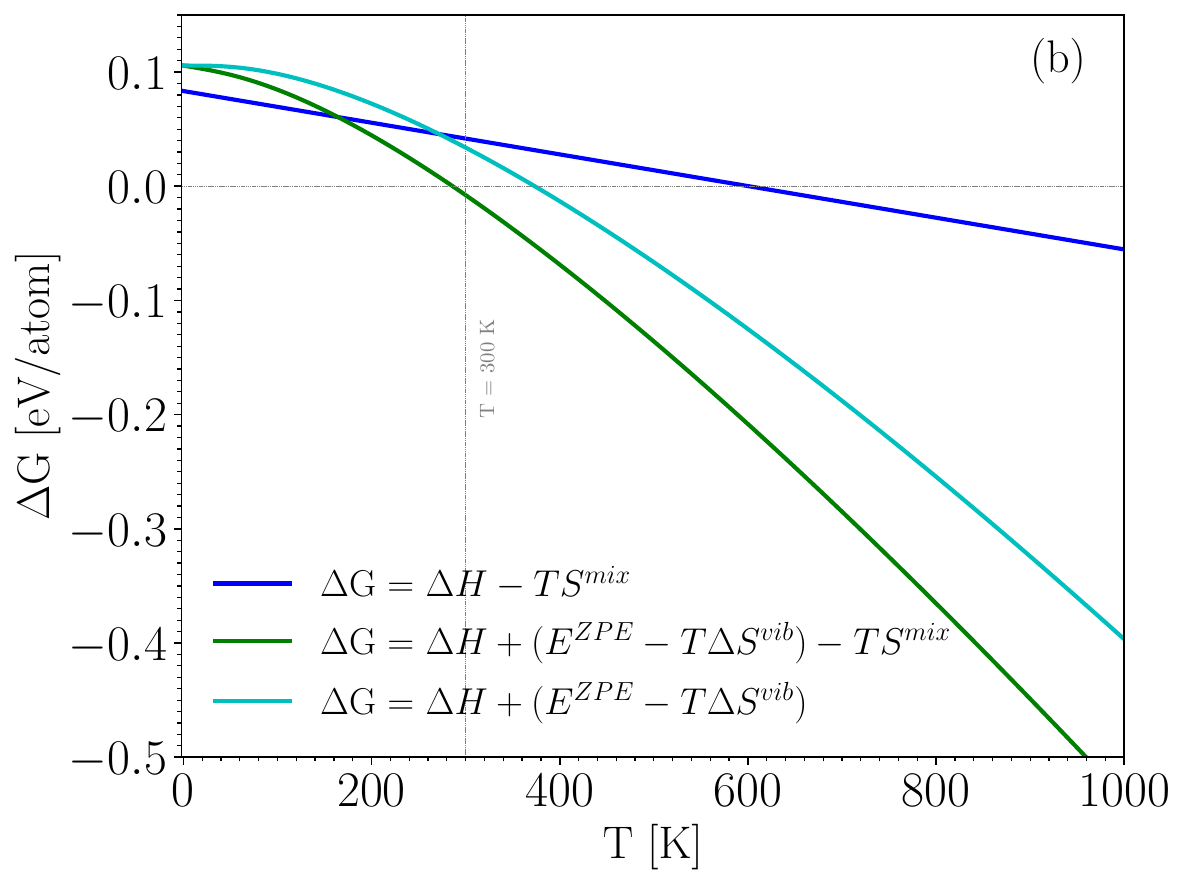}
        \caption{$~$}
    \end{subfigure}
    \caption{(a) Calculated phonon band structure of the stable structure of HfNbTaTiZr alloy. Zero reference is given by the dashed line. The acceptable lower limit criterion for numerical error is $-0.3$ THz in phonon calculations \cite{Brivio2015}. (b) Temperature dependence phase stability of HEA. }
    \label{fig:phonon}
\end{figure}

With a slightly distorted structure for the lowest energy structure using phonopy code \cite{Togo2015} 750 distorted structures were generated. From the phonon band structure as shown in Figure \ref{fig:phonon}(a) we didn't observe soft modes except around the $\Gamma$ point, which is common in phonon band structure calculations due to the remnants of stresses in the relaxed structure. Typically, a numerical error of $-0.3$ THz is acceptable \cite{Brivio2015}.

In Figure \ref{fig:phonon}(b) temperature dependence of Gibbs free energy as a function of vibrational entropic term is shown. To elaborate, we have considered three cases: First, when vibration entropic is ignored, we see the crossover temperature is around 550 K (blue line); Secondly, when the vibration entropic term is included, this crossover is lowered to 380 K (cyan line). Thirdly, when mixing energy, mixing entropy, and vibration entropy are included, the crossover temperature lowers to 280 K (green line). This implies that, at room temperature, HEA becomes dynamically stable. Thus, to model HEA we need to include the vibration entropic term. 

To further investigate whether the derivatives of our alloy also have positive mixing energy we consider four alloys NbTiZr, NbTaTi, NbTaZr, and NbTaZrTi for which experimental lattice parameters and structure stability at room temperature are reported. A recent experiment by Senkov \textit{et al.} \cite{Senkov2020} on refractory ternary equiatomic alloys such as NbTaZr and NbTaTi has also shown that these alloys have a multiphase crystal structure. Their results have shown that NbTaZr exists as a co-continuous mixture of two bcc phases. In these two phases, the elemental composition is quite different, e.g., one phase is quite rich in Nb and Ta, while the other phase is rich in Zr. In addition, a Zr-rich hcp phase was also found in the grain boundary. On the other hand, the NbTiZr alloy is reported to have a single-phase bcc structure \cite{Senkov2019, Senkov2018a}. Zhao \textit{et al.} investigated experimentally Nb-based equiatomic RHEA as-cast and reported the existence of a single bcc phase in the NbTaZrTi alloy \cite{Zhao2019}. Since the experimental data displays the bcc phase, we have modeled this system assuming bcc crystal symmetry.

\begin{table}[ht]
    \centering
    \caption{Mixing energy and Gibbs free energy of ternary, quaternary, and quinary equimolar alloys calculated from DFT and using Eq. \ref{eq:Gibbs}. The number of atoms is given for the SQS supercell generated from MCSQS. The mixing entropic term, $-T\Delta S_{conf}$, for HfNbTaTiZr, NbTaZrTi, and NbTaZr/NbTiZr/NbTaTi alloys is -41.6, -35.8, and -28.4 meV/atom, respectively. Gibbs free energy is reported at 300 K. Experimental values are reported from the references \cite{Senkov2020, Zhao2019, Zyka2019} and are given in parentheses.}
    \label{tab:T3}
    \scalebox{0.95}{
    \begin{tabular}{lccccccc}
        \toprule
         SQS alloys [this work]&	\#ofatoms&$a_{avg}$ & &$\Delta H$ &	$\Delta$G\\ 
         \cmidrule{3-3}\cmidrule{5-6}
         && \multicolumn{1}{c}{\AA} &&\multicolumn{2}{c}{meV/\AA} \\
         \midrule
            NbTiZr	    &54	    &3.387(3.396)	&&73.37	&44.97\\
            NbTaTi	    &54	    &3.293(3.296)	&&38.56	&10.16\\
            NbTaZr	    &54	    &3.399(3.354)	&&78.93	&50.53\\
            NbTaZrTi	&128	&3.361(3.359)	&&59.34	&23.54\\
            NbTaZrTiHf	&125	&3.406(3.404)	&&84.42(86)  &47.40\\
         \bottomrule
    \end{tabular}
    }
\end{table}

In particular, we would like to hypothesize the effects of the combination of hcp and bcc elements on the formation energy and also on Gibbs energy. In the absence of experimental mixing energies, we have presented the results for the SQS supercell. Ternary and quaternary supercells were generated using the same settings as for a quinary alloy, see $\mathsection2$. The mixing energies reported in Table \ref{tab:T3} were obtained by generating 20 different SQS configurations for each composition and selecting the structure with the lowest energy. The average mixing energy with standard deviation for NbTaTi, NbTaZr, and NbTiZr are $38.5\pm1.3$ meV/atom, $78.9\pm2.2$ meV/atom, and $59.3\pm3.5$ meV/atom respectively. 

From Table \ref{tab:T3}, a qualitative trend can be seen that for ternary and quaternary refractory bcc alloys, the formation energy at 0 K is slightly positive. Even when including the contribution of the configuration entropic term, the Gibbs free energy remains slightly positive. However, for the equimolar NbTaTi alloy, we observe a large difference in the decrease of the formation energy compared to the other ternary alloys. Moreover, the estimated average lattice parameters fall within the range of pure bcc Nb and Ta elements, except for NbTaZr, where Senkov \textit{et al.} reported the multiphase crystal structures. This alloy exists as a co-continuous spinodal mixture of two bcc phases and one hcp phase. The reported lattice parameters are $a_{bcc1}$ = 3.32 \AA~and $a_{bcc2}$ = 3.354 \AA~for the two bcc phases. The hcp phase has $a_{hcp}$ = 3.204 \AA $~$ and $c/a = 1.58$ \AA  \cite{Senkov2020}. 

Since our calculations are at 0 K and the alloys are synthesizable at room temperature, this entails the dynamic stability at temperature T \cite{Yao2017, DesignedResearch;J2018, Gao2017b}. Our phonon analysis demonstrated that it is necessary to incorporate vibrational entropic terms.
\subsection{Determination of elastic constants}

The elastic constants for the quinary SQS supercell were determined using the strain-energy approach. From Table \ref{tab:T2} we see there is a slight lattice distortion from bcc cubic symmetry. Therefore, we resorted to the calculation of the full elastic tensor. The resulting relaxed structure has P1 symmetry, resulting in triclinic geometry. To obtain the individual constants, a suitable choice of Lagrangian strain $(\eta_{i})_{i=1}^{6}$ was applied to the equilibrium SQS supercell. Eleven deformed structures with equal spacing in the $\delta$ = [-0.05, 0.05] were generated. The calculated energy-strain points were fitted by third-order polynomials, and then the elastic constants were derived from the second-order derivative of the energy as a function of $\eta$. The details of the calculations and deformation schemes are described in \cite{Golesorkhtabar2013}.

The elastic constant for bcc supercell was approximated with an averaging scheme given by C$_{11}$=(c$_{11}$+c$_{22}$+c$_{33}$)/3, C$_{12}$=(c$_{12}$+c$_{23}$+c$_{13}$)/3, C$_{44}$=(c$_{44}$+c$_{55}$+c$_{66}$)/3 \cite{Yin2019}. The calculated elastic constants of the SQS supercell are given in Table \ref{tab:T4}. The corresponding average values are C$_{11}$ = 153 GPa, C$_{12}$ = 104 GPa, and C$_{44}$ = 38 GPa. The values (although performed at 0 K) are in good agreement with the experiment, except for C$_{11}$ where there is a deviation of 15\%. The experimental uncertainty for C$_{11}$ and C$_{44}$ is also reported by Dirras \textit{et al.} \cite{Dirras2016} and is shown in Table \ref{tab:T4}. The experimental value for C$_{12}$ is derived from the relation B =(C$_{11}$+2C$_{12}$)/3. The theoretical EMTO-CPA study on this RHEA \cite{Fazakas2014} has also reported 160.2 GPa for C$_{11}$, 124.4 GPa for C$_{12}$ and 62.4 GPa for C$_{44}$. C$_{44}$ is largely overestimated. However, they are reported without considering the relaxation of the supercell. From the analysis, it can be seen that to obtain the elastic constant, it is necessary to consider the lattice distortion.

We checked the mechanical stability, and they all satisfy the condition, i.e., C$_{11}$ > C$_{12}$, C$_{11}$ + 2C$_{12}$ > 0 and C$_{44}$ > 0. Two constants E and G were approximated using the Hills approximation to characterize the stiffness of the material \cite{Hill1952}. The shear modulus is quite low compared to the constituent elements. The bulk modulus is 10\% lower than the experimental value. The elastic modulus (E) is calculated using Hill's approximation (see Appendix). The Pugh ratio B/G is usually used to estimate the brittle-ductile nature of the material \cite{Pugh1954}. A critical value of 1.75 is used to distinguish brittle material from ductile material. A higher ratio indicates ductility, while a lower ratio indicates the brittleness of the materials. In our calculation, the ratio is higher and follows the experimental trend.

\begin{table}[ht]
    \centering
    \caption{Elastic constants C$_{ij}$ (GPA), bulk modulus B (GPA), shear modulus G (GPA), Young's modulus E (GPA), and Pugh ratio B/G as computed using the SQS supercell for HfNbTaTiZr alloy.}
    \label{tab:T4}
    \begin{tabular}{lccccccc}
        \toprule
         SQS	&C$_{11}$	&C$_{44}$	&C$_{12}$  &B	    &G	    &E	    &B/G\\ 
         \midrule
        125	 &153	    &38	    &104	   &121	&32.0	&88	   &3.8\\
        EXP\cite{Dirras2016}	&172±6	   &28±1.5	    &108	   &135	&28	    &79	&4.3\\
         \bottomrule
    \end{tabular}
\end{table}


\section{Core structure of the screw dislocations}

In the literature, the core structure of screw dislocations for pure bcc elements has been extensively studied over the last two decades. To study the core structure of $\frac{1}{2}\langle111\rangle$ screw dislocations for HEA, the local composition in different regions of the supercell must be explored. For pure systems, many authors have studied the size, geometry, and arrangement of screw dislocations in a supercell by studying the difference between easy and hard-core energies within trigonal, dipole, and quadrupolar arrangements and selecting the optimal parameters for studying the core structure  \cite{Lib,Cai2005,Clouet2009,Ventelon2007}. Further, the issue of conditional convergence of long-range elastic fields in a finite simulation box where PBC conditions are used. However, the regularization method proposed by Cai \textit{et al.} \cite{Caiyz2003} addresses this problem. Clouet \cite{Clouet2011a} has shown that the elastic energy can be estimated by considering the large periodic images of the supercell within the anisotropic linear elasticity theory. This approach has been validated for simple pure systems, but for inhomogeneous systems, the question of the reliability of the calculation of the core energy arises. Moreover, to compute the dislocation core energies \cite{Clouet2011b} for bcc Fe, Mo, and Ta, the dislocation core fields in addition to the Volterra field have been considered when screw dislocations are inserted. 

For pure systems, homogeneous elastic fields are observed. However, for HEAs, local chemical fluctuations due to random atoms produce inhomogeneity at each atomic site. Therefore, we would expect that there is always a variation of the elastic strain with distance and the contribution of core fields. Currently, no methodologies exist in the literature to consider these local fluctuations hence as an approximation we are relying on our results on the formalism of pure systems.

\begin{figure}[ht]
    \centering
    \includegraphics[scale=0.27]{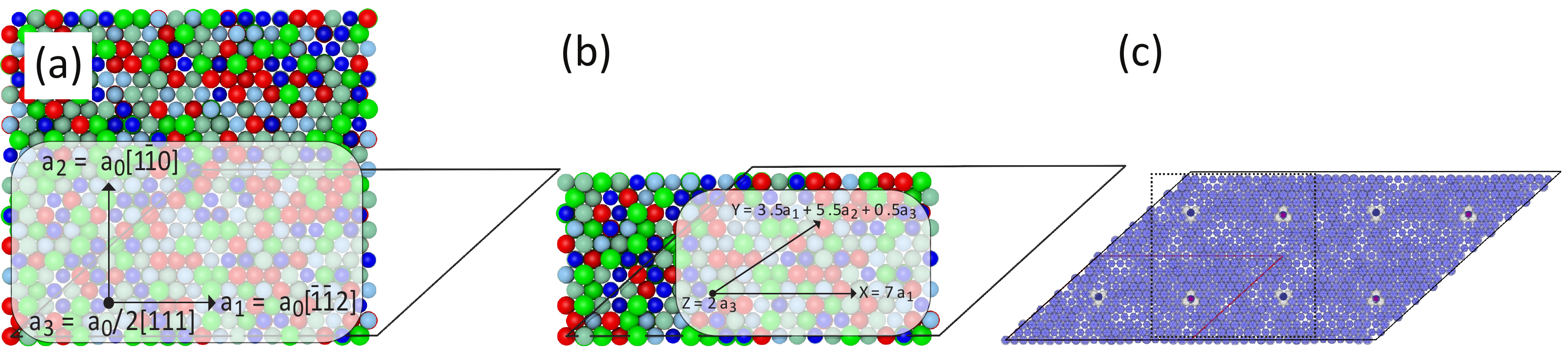}
    \caption{Construction of a dislocation dipole cell (b) by transforming the dislocation quadrupole cell (a) using the approach described in detail by Li \textit{et al.} \cite{Lib}. A common neighbor analysis is used to visualize the core localization in (c). The blue color represents the bcc environment, while the white color represents non-bcc (screw dislocations). }
    \label{fig:F2}
\end{figure}

To calculate the core energy, we have used the quadrupole dislocation geometry. The advantage of this geometry is the rapid convergence of easy and hard-core energies in contrast to other dislocation arrangements (\textit{see} ref \cite{Clouet2009}). In addition, the quadrupole arrangement approach has been validated by Weinberger \textit{et al.} and Ventelon \textit{et al.} \cite{Ventelon2007,Weinberger2013b} for pure bcc systems. Figure \ref{fig:F2}(b) shows the construction of the tilted periodic supercell containing 462 atoms. To obtain the relaxed core structure, in contrast to pure systems, for RHEA, we proceeded as follows: First, the SQS was generated for the given supercell and then the atomic position was relaxed to the equilibrium ground state structure. Then, a dislocation dipole was introduced into the simulation cell, followed by relaxation of the structure again. The screw was introduced using the anisotropic displacement field from \cite{Clouet2009} to help the system converge faster.

When studying dislocations in the special case of RHEA, additional constraints arise, such as the number of layers to be considered in the Z-direction. On the one hand, the dimension of the $z-$axis (corresponding to the screw dislocation line) must be large enough to sample the chemistry in this direction and limit short-range correlations induced using PBC. However, our preliminary results showed that different core structures could be observed depending on the local chemistry adjacent to the dislocation core in the present system. This was also observed for a ternary bcc alloy \cite{Rao2019}. On the other hand, if the $z$ supercell dimension is relatively large, multiple core structures can be observed along the same dislocation line. The presence of different core structures along the same dislocation line prevents accurate measurement of the core energy or its correlation with local chemistry. Therefore, our simulations dealing with dislocations include 2\textbf{b} of thickness, and we ensured that the core structure along each of the dislocations was the same.

The core energy, E$_{core}$, of a screw dislocation in a dipole supercell is estimated using the relation derived by Clouet \textit{et al.} \cite{Clouet2009} (\textit{see} Eq. \ref{eq:eq4}). In this relation, the excess energy E is obtained by taking the energy difference with and without the dislocation dipole from DFT. On the right-hand side of the equation, the contribution of elastic energy, $E_{Elastic}$, is calculated using the anisotropic elastic theory, for which we have used the elastic constants calculated in Section 3.2.

\begin{equation} \label{eq:eq4}
     \mathrm{E} = 2\mathrm{E}_{core} + \mathrm{E}_{elastic}
\end{equation}

Before analyzing our system, we first validated this approach on the pure bcc Ta system \cite{Weinberger2013b} to ensure that the implementation of the anisotropic displacement field and core analysis work as expected. The core structure of our RHEA was studied for all the different locations of the dipole corresponding to different local chemistries. Unlike results on Ta or existing data on pure bcc elements, the core structure is not always found to be compact but sometimes close to a degenerate or split structure, depending on the local chemistry. Typically, in most simulations, one core would tend to be compact and the other non-compact. In addition, the observed compact core does not correspond exactly to the textbook compact core structure, which has a 3-fold symmetry.

To analyze the core structure in HEA, one must consider the choice of reference structure in mapping the displacement along the $z$-axis. To rationalize this, we have chosen the relaxed HEA structure without the dislocations (but relaxed) as a reference. The displacement mapping does not yield more distinct compact core structures than when using the perfect bcc supercell.

To set these results in perspective for our alloy, in all configurations containing a dipole, we observed that one core structure is almost compact, and the other corresponds to a split or to another structure. This variation in the core structure has been observed for ternary alloys by Rao \textit{et al.} \cite{Rao2019}. On the other hand, in HfNbTaTiZr alloy, one core is almost compact, and the other is split in the $\{110\}$ plane. As mentioned earlier, HfNbTaTiZr is reported to have a single bcc phase with no multiphase crystal structure \cite{Zhao2019}. The observed variation in core structure could be due to the asymmetry of local composition at the core sites, leading to the variation along the dislocation line.


\begin{figure}[ht]
    \centering
    \includegraphics[scale=0.6]{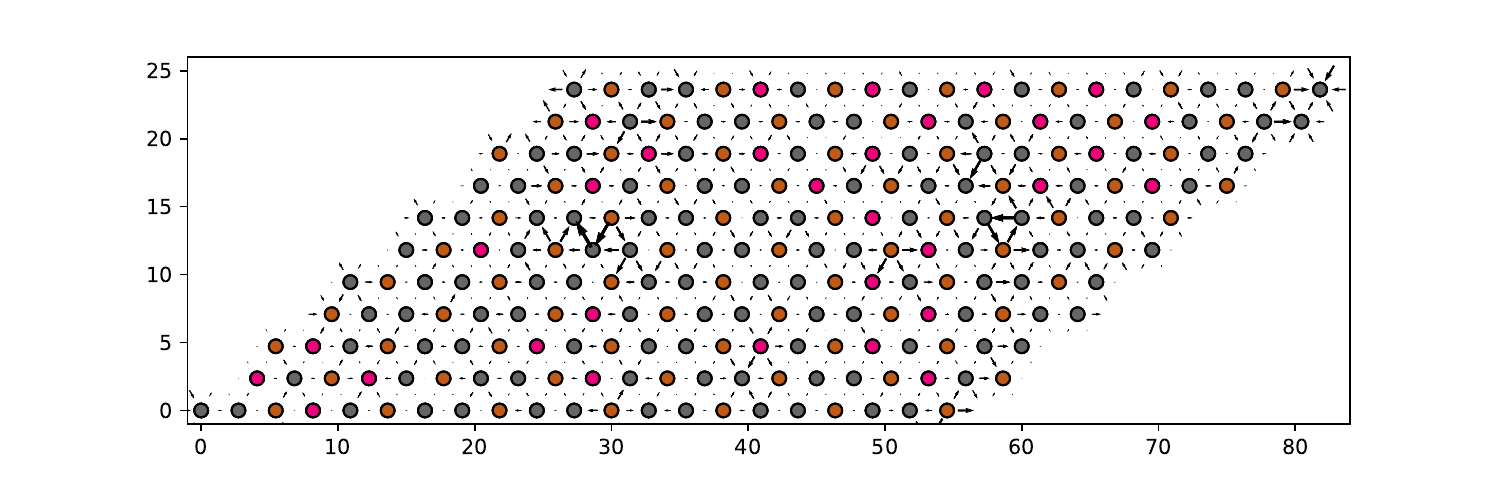}
    \caption{Compact and split core structure of $\frac{1}{2}\langle111\rangle$ screw dislocations in HfNbTaTiZr alloy visualized with the differential displacement map \cite{Hale2017}. The colored dots represent the atoms in the different layers. The bold vectors represent the magnitude of the displacement.}
    \label{fig:F3}
\end{figure}

\section{Conclusions}
We investigated the structure, stability, and elastic properties of bcc RHEA HfNbTaTiZr using DFT. The SQS supercell, consisting of 125 atoms, can reproduce the HEA randomness while preserving the bcc structure. When modeling random alloys, multi-body correlations do not lead to much improvement. The lattice parameters and elastic constants are in good agreement with experimental data. The Gibbs formation energy is slightly positive for quinary HEA. This suggests that a contribution of short-range chemical ordering may be required to stabilize the RHEA solid solution or the existence of local phases. The core structures are often compact but also split. This slight variation is different from the purest bcc structure.

\section{Acknowledgements}
We are grateful to Flemming Ehlers for his suggestions in preparing the manuscript. We are also grateful to Univ. Sorbonne, Paris Nord, Labex SEAM, ANR (Agence Nationale de la Recherche), and CGI (Commissariat à l’Investissement d’Avenir) for financing this work through Labex SEAM (Science and Engineering for Advanced Materials and devices), ANR-10-LABX-0096 and ANR-18-IDEX-0001. Further, we would like to thank Ing. Nicolas Greneche for HPC support, MAGI, Univ. Sorbonne Paris Nord and HPC resources from GENCI-[CCRT/CINES/IDRIS] (Grant 2019[A0060807006]).

\section*{Appendix}
\section*{Determination of mechanical properties}

The mechanical properties were determined via the elastic constants. The Voigt and Reuss averages for bulk modulus B and shear modulus G are:

\begin{equation} 
    B_{V}=(C_{11}+C_{22}+C_{33}+2(C_{12}+C_{23}+C_{13}))/9
\end{equation}

\begin{equation}
    B_{R}=1/(S_{11}+S_{22}+S_{33}+2(S_{12}+S_{23}+S_{13}))
\end{equation}

\begin{equation}
    G_{V}=(C_{11}+C_{22}+C_{33})-(C_{12}+C_{13}+C_{23})+3(C_{44}+C_{55}+C_{66}))/15
\end{equation}

\begin{equation}
    G_{R}=15/(4(S_{11}+S_{22}+S_{33})-4(S_{12}+S_{13}+S_{23})+3(S_{44}+S_{55}+S_{66}))
\end{equation}

where S is a compliance matrix. $G_{R}$ and $G_{V}$ represent the lower (Reuss) and upper (Voigt) limits, respectively. The shear moduli G and B are obtained by taking the average $(G_{R}+G_{V})/2$ and $(B_{R}+B_{V})/2$, respectively, as suggested by Hill \cite{Hill1952}. The Young’s modulus, E, and Poisson's ratio, $\nu$, can be obtained using B and G,

\begin{equation}
    E = 9BG/(3B+G);
    \nu = (3B+2G)/2(3B+G)
\end{equation}

To evaluate the reliability of the elastic constants, the two parameters Zener anisotropy ratio $A_{z}$ and Voigt-Reuss elastic anisotropy $A_{GVR}$ are used to characterize the elastic anisotropy of the material. They are,

\begin{equation}
    A_{Z}=2C_{44}/(C_{11}-C_{12});
    A_{GVR}=(G_{V}-G_{R})/(G_{V}+GR)
\end{equation}

For elastically isotropic material, the $A_{z}$ ratio goes to one and $A_{GVR}$ zero.

\clearpage
\bibliographystyle{ieeetr}
\bibliography{library.bib}
\end{document}